\documentstyle{caosp}
\begin{document}
\pubyear{1998}
\volume{27}

\firstpage{353}
\hauthor{L. \v C. Popovi\'c and M. S. Dimitrijevi\'c}

\htitle{ The electron-impact
broadening parameters for ionized  RRE  lines}

\title{ A programme for electron-impact
broadening parameter calculations
of ionized rare-earth element lines}

\author{L. \v C. Popovi\'c and M. S. Dimitrijevi\'c}

\institute{Astronomical Observatory, Volgina 7, 11000 Belgrade,
Serbia, Yugoslavia
E-mail: lpopovic@aob.bg.ac.yu; mdimitrijevic@aob.bg.ac.yu}

\date{November 10, 1997}

\maketitle

\begin{abstract} In order to provide atomic data needed for astrophysical
investigations, a set of electron-impact broadening parameters for  ionized
rare-earth element lines should be calculated.  We are going to
calculate the
electron-impact broadening parameters for more than 50 transitions of ionized
rare-earth elements. Taking into account that the spectra of these
elements are
very complex, for calculation we can use the modified semiempirical
approach -- MSE or simplified MSE. Also, we can estimate these
parameters on the basis of regularities and systematic trends.


\end{abstract}

\section{Motivation}

The spectral lines of rare-earth elements are present in Solar as well as in
stellar spectra (see e.g.  Grevesse \& Blanquet 1969, Molnar 1972,
 Adelman 1987, Mathys \& Cowley 1992, Sadakane 1993, Bidelman et al. 1995,
Cowley et al. 1996, etc.). Principally, these lines  originate in layers
 of stellar atmospheres with higher electron density (photosphere or
subphotosphere). Consequently, electron-impact broadening mechanism can be
important, especially for hot (A and  B) stars as well as for
white dwarfs. So, it is important to have a set of electron-impact
broadening data for
the lines of ionized rare-earth elements. For some transitions of La\,{\sc ii}
and La\,{\sc iii} we have calculated Stark widths
(Popovi\'c \& Dimitrijevi\'c 1997) by
using the modified  semiempirical approach (Dimitrijevi\'c \& Konjevi\'c 1980,
Popovi\'c \& Dimitrijevi\'c 1996a,b). Here we  present our plans and specify
the number of lines for which we may calculate electron-impact broadening
parameters with a satisfying accuracy and discuss the difficulties which may
appear in the calculation.

\section{\bf Methods of calculation}

Due to the lack of known energy levels as well as of reliable transition 
probabilities for rare-earth elements, the approximate methods are adequate for
Stark broadening calculations. Consequently the modified semiempirical approach
will be applied. This method was developed by
Dimitrijevi\'c \& Konjevi\'c(1980). For
the case of ions with complex spectra the improvement was done by Popovi\'c \&
Dimitrijevi\'c (1996a,b). Also, as regards lines for which it is not possible
to apply this method, we will use the simplified modified semiempirical formula 
(SMSE) given by Dimitrijevi\'c \& Konjevi\'c (1987). For the lines which are 
very important for astrophysical purposes and for which, due to the lack of 
atomic data, it is not possible to use  even the SMSE method, we will estimate 
Stark broadening parameters on the basis of regularities and systematic trends 
(RST, Dimitrijevi\'c \& Popovi\'c 1989).

\begin{table}[ht]
\small
\begin{center}

\caption{
List of the ions for which we are going to calculate the electron-impact
broadening parameters. The number of transitions given in  the table could be
calculated using the modified-semiempirical (MSE) and simplified
modified-semiempirical (SMSE) methods, the $x$ indicates that the data can be 
provided for several other transitions by using regularities and systematic 
trends (RST) for astrophysically very important lines. Key to the columns:
I, IV -- Ion, II, V -- Number of transition for which we can calculate the
Stark broadening parameters, III, VI -- Method which we are going to use.}

\label{t1}
\begin{tabular}{llllllllll} \hline\hline

 \ \ \ I &\ \ \  II  &\ \ \  III &&&\ \ \  IV &\ \ \  V  &\ \ \  VI && \\
\hline

 La II & 3+x  & SMSE+RST  &&& La III & 6+x & MSE+RST && \\
 La IV &\ \ \ \  x & RST &&& Ce II   &\ \ \ \  x   & RST && \\
 Ce III  & 5+x   & SMSE+RST &&& Ce IV   & 4+x   & MSE+RST  && \\
 Pr II,III  &\ \ \ \  x   & RST &&& Nd II  & 5+x   & SMSE+RST  && \\
 Nd III  &\ \ \ \   x  & RST &&& Sm II  &\ \ \ \   x  & RST  && \\
 Eu III  & 2+x   & SMSE+RST &&& Gd II  & 2+x   & SMSE+RST  && \\
 Tb III  & 3+x   & SMSE+RST &&& Ho II   & 2+x   & SMSE+RST  && \\
 Ho III  &\ \ \ \  x   & RST &&& Er II   & 1+x   & SMSE+RST  && \\
 Er III  &\ \ \ \  x   & RST &&& Tm II,III   &\ \ \ \   x  & RST  && \\
 Yb II  & 5+x   & MSE(SMSE)+RST &&& Yb III  & 3+x   & MSE+RST  && \\
 Yb IV  &\ \ \ \   x  & RST &&& Lu II  & 2+x   & SMSE+RST  && \\
 Lu III  & 5+x   & MSE+RST &&& Lu IV  & 3+x   & MSE+RST  && \\
\hline\hline
\end{tabular}
\end{center}
\end{table}

Moreover, due to the very complex spectra of ionized rare-earth elements we
have to improve the existing software developed by us (Popovi\'c 1994). It
means that calculations within intercoupling approximation have to be
performed. For example in the spectra of  Ce\,{\sc iii}, the $4f6p$ levels are 
well described by $jj$ coupling approximation, while $4f6d$ levels, which are
perturbed by  $4f6p$ ones, are well described by $j\ell$ coupling
approximation. Such interaction between these two levels should be taken into
account. Also, a numerical experiment about the influence of this effect on
calculated parameters should be done. \medskip

\section{The list of ions}

In Table 1 we present the ions and number of lines for  which we are going to
calculate the electron-impact broadening parameters. As one can see from
Table 1, there is a very limited number of transitions for which this is 
possible (only 51 transitions). The list has been made
taking into account atomic data given by Martin et al. (1978), so, this list 
may be extended after a detailed search through literature and after including 
the new experimental results.

{}
\end{document}